\documentclass[a4paper, 11pt]{article}

\usepackage[margin=1in]{geometry}
\usepackage[utf8]{inputenc}
\usepackage{authblk}
\usepackage{microtype}
\usepackage[T1]{fontenc}
\usepackage{xurl}
\usepackage{hyperref}
\usepackage[hyphenbreaks]{breakurl}
\usepackage{tabularx}
\usepackage{booktabs}
\usepackage[flushleft]{threeparttable}
\usepackage[singlelinecheck=off,font=small,labelfont=bf]{caption}

\usepackage{fancyhdr}
\title{AI Certification: Advancing Ethical Practice by Reducing Information Asymmetries}

\author[1]{Peter Cihon}
\author[2]{Moritz J. Kleinaltenkamp}
\author[3]{Jonas Schuett}
\author[4]{Seth D. Baum}
\affil[1]{\small \emph{GitHub}}
\affil[2]{\small \emph{Hertie School, Centre for Digital Governance}}
\affil[3]{\small \emph{Legal Priorities Project}}
\affil[4]{\small \emph{Global Catastrophic Risk Institute}}
\date{\vspace*{-1em} \small Forthcoming, \emph{IEEE Transactions on Technology and Society}, \href{https://doi.org/10.1109/TTS.2021.3077595}{doi:10.1109/TTS.2021.3077595} \\ This version 2 May 2021}

\begin{document}
\maketitle
\vspace*{-1.5em}
\thispagestyle{fancy}

\begin{abstract}
\noindent As artificial intelligence (AI) systems are increasingly deployed, principles for ethical AI are also proliferating. Certification offers a method to both incentivize adoption of these principles and substantiate that they have been implemented in practice. This paper draws from management literature on certification and reviews current AI certification programs and proposals. Successful programs rely on both emerging technical methods and specific design considerations. In order to avoid two common failures of certification, program designs should ensure that the symbol of the certification is substantially implemented in practice and that the program achieves its stated goals. The review indicates that the field currently focuses on self-certification and third-party certification of systems, individuals, and organizations—to the exclusion of process management certifications. Additionally, the paper considers prospects for future AI certification programs. Ongoing changes in AI technology suggest that AI certification regimes should be designed to emphasize governance criteria of enduring value, such as ethics training for AI developers, and to adjust technical criteria as the technology changes. Overall, certification can play a valuable mix in the portfolio of AI governance tools.
\end{abstract}

\section{Introduction}
\label{1}

The growing societal implications of AI have prompted extensive attention to the ethics principles that should guide AI development and use \cite{1,2} and, more recently, some attention to how organizations can translate these principles into practice \cite{3,4,5}. These are vital areas of activity, but they leave open an important question: how do external stakeholders know whether the organizations and their AI systems are indeed meeting their ethical standards? This paper discusses one potential answer to this question: certification.

Certification can be defined as the attestation that a product, process, person, or organization meets specified criteria (paraphrased from \cite{6}). Certification involves an assessment of the entity to be certified, often, but not necessarily, by a trusted third party. If the assessment determines that the entity meets the specified criteria, then the entity is certified accordingly. Thus, the certification process reveals information about the inner workings of the entity that can be provided to external stakeholders, reducing information asymmetries between insiders and outsiders \cite{7}. In doing so, certification incentivizes insiders to adhere to higher standards because they can get credit for it if they do and get caught if they do not. In short, certification is a governance tool for advancing transparency and incentivizing insiders to meet specified criteria. Those criteria could include voluntary AI ethics principles or mandatory regulatory requirements.

Certification is already in wide use in many sectors \cite{8,9} and is starting to appear for AI. Certification includes such disparate phenomena as the US EnergyStar program for consumer appliance energy efficiency, ISO 9001 for quality management systems in global supply chains, and the mandatory ``CE mark'' for various products sold in the EU. Certification programs for AI systems have been proposed or are currently being developed by the European Commission \cite{10}, the IEEE \cite{11} and by the Chief Scientist of Australia \cite{12}; one has already launched from the government of Malta \cite{13}. Education certification programs include the Queen's University executive education program on Principles of AI Implementation \cite{14} and the Finnish civic education program Elements of AI \cite{15}. Additionally, a coalition of Danish organizations \cite{16} is developing an IT security certification for organizations that includes AI-related criteria. These various programs speak to the diverse forms that certification programs can take, including public and private, voluntary and compulsory, and for the AI systems themselves as well as the people and organizations involved with them.

This paper presents an overview of certification for advancing ethical AI. It provides background on AI certification, drawing on management literature (Section \ref{2}), surveys current AI certification programs and proposals for new programs (Section \ref{3}), and discusses prospects for certification to address issues raised by future AI technologies (Section \ref{4}). Section \ref{5} concludes.

\pagenumbering{arabic}
\setcounter{page}{2}

\section{Background}
\label{2}

This section reviews management literature on the merits of certification as it relates to the governance of AI. This literature typically focuses on market behavior of corporations and their customers. However, certification is also relevant to nonmarket settings. For example, AI systems developed by nonprofit and/or open source teams could seek certification to build user confidence, even if users are not charged for using the systems. Or, governments might seek international certification of their military AI systems to demonstrate compliance with international humanitarian law or other standards. Thus, while certification is important for the corporate governance of AI (on which, see \cite{17}), it is also more widely applicable.

First, some terminology. The \emph{object of certification} is the entity to be certified. It may be a product, process, individual, or organization. The \emph{certifier} is the actor who does the certifying. The \emph{assessor} is the actor who assesses whether or not an object of certification meets the specified criteria prior to certification. The assessor and certifier can be one and the same or they can be two separate actors, such as in the case of certification programs that use third-party assessment. The \emph{assessment} or \emph{conformity assessment} is the evaluation of whether the object meets the specified criteria; this can vary in stringency, from a review of application materials to an audit of systems or processes. The \emph{applicant} is the actor who controls the object of certification and seeks to obtain certification when it is voluntary. The applicant and object of certification can be the same, such as in the case of skills certifications granted to individuals. Alternatively, the applicant can be the entity who oversees, manages, develops, or otherwise controls the object of certification. Finally, the \emph{audience} is the group of people who will be informed by the certification. For example, consider an AI system developed by a corporation, certified by a government agency, assessed by an independent consulting firm, to inform consumer purchasing decisions. Here, the object is the AI system, the certifier is the government agency, the assessor is the consulting firm, the applicant is the corporation, and the audience is consumers. In some cases, the applicant may also simultaneously be the certifier; this is known as self-certification. Self-certification creates conflict of interest, though this can sometimes be addressed via measures such as audits and liability regimes.

Two additional sets of terms help discuss the governance potential of certification programs: First, \emph{symbol-substance coupling} refers to alignment between the statement that an object meets the certification criteria (the symbol) and that object actually meeting those criteria (the substance). Likewise, symbol-substance \emph{decoupling} is when a certification inaccurately characterizes its object \cite{18,19,20}. Second, \emph{means-ends coupling} refers to alignment between the means of the certification program (i.e., the specific criteria to which objects are being held) and the ends underlying its design \cite{19,21}. Means-ends \emph{decoupling} is when a certification program fails to advance its intended goals. Symbol-substance decoupling and means-end decoupling are two failure modes of certification programs.

\begin{table}
\caption{Summary of Key Certification Concepts}
\label{table_1}
\scriptsize
\begin{tabularx}{\textwidth}{ p{4cm} X X }
	\toprule
	\textbf{Concept} & \textbf{Definition} & \textbf{AI Examples} \\
	\midrule
	Object of certification & Entity to be certified, in order to reduce information asymmetries & AI systems, AI development processes, AI engineers, AI corporations \\ 
	Certifier & Actor doing the certifying & Governments, non-profits, corporations \\ 
	Assessor & Actor assessing whether the object of certification meets specified criteria & Governments, non-profits, corporations \\ 
	Applicant & Actor controlling the object of certification & AI engineers, AI corporations \\ 
	Audience & Actors benefiting from the reduction of information asymmetries & Market actors engaging with the object of AI certification in question, regulators \\ 
	Symbol-substance decoupling & The state in which certification programs fail to accurately characterize their objects of certification such that false positives and false negatives arise & ``Ethics washing'' by AI corporations, in which they proclaim to be holding themselves to ethics principles but do not actually enforce them in their practices, services, and products \\ 
	Means-ends decoupling & The state in which certification criteria, even if properly enforced, do not actually bring about the changes intended by the designers of the certification program & AI certification programs that purport to address ethics broadly but enforce overly simplistic, abstract, or restrictive practices, thereby inhibiting more effective responses \\ 
	Credence attributes & Attributes of products and services that can be identified by outside actors only after repeat observations and/or at great cost & Bias, explainability of AI systems \\ 
	Potemkin attributes & Attributes of products and services that cannot be identified by outside actors at all; they require insider information to be identified & Whether an AI system was developed according to ethics principles, its security, the extensiveness of testing it underwent \\ 
	\bottomrule
\end{tabularx}
\end{table}

\subsection{Reducing Information Asymmetries}
\label{2.1}

As noted above, a primary role of certification is to reduce information asymmetries. Information asymmetries raise two issues: the ``selection problem'' of identifying desirable suppliers of a good or service, and the ``monitoring problem'' of evaluating whether the supplier is meeting specified agreements \cite{22}. Both problems create space for rent-seeking behavior, in which the supplier extracts greater wealth from other parties without adding social value \cite{23}. Information asymmetries are acute in AI systems because the systems are often complex and opaque and users typically lack the data and expertise necessary to understand them.

Certification is especially well-suited to addressing such information asymmetries that arise from ``credence'' and ``Potemkin'' attributes of goods and services \cite{24}. \emph{Credence attributes} are those that can be identified by outside parties, but only as trends that emerge after repeat observations. 

Individual audience members may struggle to identify credence attributes on their own, but an organized third party, such as a certification program, can collect the data needed for a reliable characterization. Credence attributes of AI systems include whether a system is fair or unbiased, is explainable across a wide range of use cases, protects user privacy across a wide range of settings, and is safe and secure to rare and/or diverse threats. \emph{Potemkin attributes} present even greater information asymmetry challenges: they cannot be identified at all by outside parties, and instead require insider information. An important type of Potemkin attribute is the process through which the good or service was made, such as whether an AI system was developed in accordance with labor, environmental, or ethical standards, or whether it was rigorously tested prior to deployment. Certification programs can address both credence and Potemkin attributes, as seen for example in the popular ISO 9001 \cite{25} certification for organization quality management and ISO 14001 \cite{26} certification for environmental management. By providing information about credence and Potemkin attributes, suppliers can create new signals to attract outside parties, including market customers \cite{27} and others. Table \ref{table_1} presents a summary overview of the key concepts underlying our understanding of certification, and how we apply them to the field of AI.

A certification program reduces information asymmetries when it accurately characterizes the objects that it certifies. It must avoid false positives, in which an object is certified even though it fails to meet the specified criteria, and false negatives, in which an object meets the criteria but is denied certification. Accurate certification thus exhibits high symbol-substance coupling. 

Research on certification has identified strategies for improving certification accuracy, i.e., improving symbol-substance coupling. First, the objects to be certified should be assessed continuously, rather than only once or not at all \cite{7,28,29,30}. This ensures that the information conveyed by the certification stays up to date. Second, assessments should be made by qualified independent assessors using their own evidence and not just documentation provided by the applicants \cite{31,32,24}. This is needed to protect against applicants misleading the assessor, especially to gain a false positive. These strategies are of further value for incentivizing good behavior by the object and/or applicant, as discussed in the section below. Indeed, the strategies were largely developed to promote behavior change.

To achieve accuracy, AI certification programs will often need to parse technical details of the systems and their development. Evaluation methods required for accurate AI certification are at an early stage of development. The most mature area may be documentation for AI system training data and model characteristics \cite{33,34,35}, which can help verify certification criteria associated with both credence and Potemkin attributes. Supply-chain and development documentation methods that draw on insights from other fields may help verify Potemkin-attribute criteria \cite{36,37}. Further, benchmark tests to assess AI system performance for specific tasks in specific settings \cite{38} and more general behavioral settings for specific types of systems \cite{39} can help verify credence-attribute criteria. However, methods for criteria verification in broad areas of fairness, transparency (explainability), safety, and security remain under development. Certification programs aiming to address these areas will need to update their criteria over time as the state of the art advances.

\subsection{Incentivizing Change}
\label{2.2}

In reducing information asymmetries, certification can further serve to incentivize changes to applicant practices. For example, research has shown that corporations may be more motivated to achieve ethics standards if they can use certification to demonstrate their achievements to customers who value these achievements \cite{31,40}.

Where there is demand for certification—e.g., from corporate customers who value ethics achievements—certification programs can succeed on a voluntary basis, with no government regulation \cite{9}. Government regulation can further boost voluntary programs via tort liability \cite{41} or tax credits. Mandatory certification programs display higher compliance levels, though voluntary programs are sometimes found to be more cost-effective \cite{42}.
For a certification program to change applicant practices according to its specified criteria, it is important that it accurately characterizes these practices and their outputs. In particular, false positives permit applicants to pursue certification without changing their underlying practices: they can adopt the certification criteria symbolically but not substantively. Such symbolic adoption is of particular concern in AI. The recent popularity of AI ethics principles has prompted concerns about ``ethics washing'' \cite{43,44}, in which AI groups articulate ethics goals to bolster their reputation but do not act accordingly. An accurate certification regime could mitigate the problem of AI ethics washing.

Prior research identifies several strategies to motivate substantive instead of symbolic participation in certification programs. First are strategies to improve certification accuracy as discussed above. Second, certification assessments should not be onerous on the applicants; otherwise, applicants that meet the assessment criteria may decline to participate in the program \cite{24}. Third, the cost to applicants of failing an assessment should be high (in terms of damaged reputation, lost market power, etc.), so as to incentivize participating applicants to meet the specified criteria \cite{24,31,27,22}. A successful certification program may need to promote its value so that other parties (e.g., customers) know to penalize applicants that fail certification assessments. Finally, certification programs should require applicants to integrate the values and norms underlying the criteria into their organizational identities and strategies; this makes substantive adoption more likely \cite{30,45}.

A robust body of research suggests that properly constructed certification programs can couple symbol and substance to accurately characterize and induce changes to corporate practices. For example, studies of ISO 14001 \cite{26} have found a significant correlation between (1) the ISO 14001 attestation and the presence of environmental management practices in corporations, as well as (2) the presence of these practices in corporations and their improved environmental performance \cite{7,46}. Similar evidence exists for the ISO 9000 series on quality management practices \cite{47,49}.

Ultimately, adherence to certification criteria depends on applicants’ actions. Their actions can derive from characteristics of themselves in addition to characteristics of the certification program. Indeed, applicants sometimes engage in a mix of symbolic and substantive adoption even within the same certification program \cite{47}. This may be due to a lack of rational reflection or to internal disagreement within the applicant about the importance of the certification criteria \cite{45}. Additionally, some studies have found that firms with higher revenues tend to adopt certification more substantively because they have the resources to meet the certification criteria \cite{51,52}. Therefore, while it is important for certification programs to be well designed, the applicants retain an essential role.

\subsection{Ensuring Certification Advances the Right Goals}
\label{2.3}

The preceding section discussed how to ensure that certification criteria are in practice actually met, i.e., that certification is adopted substantively and not just symbolically. The current section covers the selection of the criteria themselves. These criteria should be designed such that if they are achieved, the certification program will deliver improvements on its underlying motivation: advancing ethics principles, making progress on societal issues, etc. In other words, the means of certification must be coupled to its ends. Means-ends decoupling can arise when developers of certification programs fail to treat the focal problem holistically, leading them to misidentify the causal relationships driving the issue at hand and prescribe the wrong action for a particular problem \cite{21}. This can spur at least three types of mistakes.

First, certification may be inadequately customized for the circumstances it addresses. A common shortcoming of certification programs is the use of uniform, homogeneous prescriptions that neglect important local and time-specific factors \cite{21}. For example, AI systems required to use a certain dataset that performs well on demographic diversity could end up underperforming when even better datasets become available. Some of the most popular corporate certification programs make homogeneous prescriptions for corporations across the globe, though a context-dependent multiplicity of corporate practices may be necessary to address societal problems across environments. Certification criteria that are customized for particular circumstances may tend to perform better, though the development of such criteria may require more resources.

Second, certification can cause unintended harms. Narrowly crafted certification criteria can neglect important causal relationships; when followed, they may make some things better but other things worse. For example, AI systems meeting certain performance standards may require additional computing power, which inadvertently harms the environment via increased energy consumption \cite{53}. This type of problem is also known as the ``waterbed effect'' because lying on a waterbed simply displaces the water: the problem ``rises'' again somewhere else \cite{21,54}. To avoid unintended harms, certification programs have to avoid overly simplistic prescriptions and instead embody and promote systemic thinking about the issues at hand \cite{21}.

Third, certification criteria can be overly restrictive, inhibiting applicants from taking innovative actions that would better address the particular problem. One way to make actions more customized to complex causal relationships expressed in particular circumstances is to permit applicants to make decisions on a case-by-case basis. However, certification criteria often prescribe a rigid set of actions, thereby inhibiting applicants from identifying better solutions \cite{21,55}. Certification programs can mitigate this type of problem by “stimulating internalization”, meaning that they de-emphasize rigid rules and instead emphasize changes to applicants’ internal culture, such that applicants are likely to do the right thing as circumstances dictate \cite{21}.

Related to these challenges is the question of whether a certification program addresses the right focal problem(s) to begin with. Exactly which problems should be focused on is ultimately a question of ethics. Opinions may vary on which issues are most important. For example, many AI experts are divided on whether the field should focus on issues raised by near-term or long-term AI \cite{56}. Certification programs may sometimes need to take sides on these sorts of divides, such that they ``cannot please everyone'', with some contending that the programs work on the wrong issues. In other cases, certification programs could emphasize issues that may seem appealing at first glance because of their broad scope, but become disagreeable when actors’ interpretations of this broad scope begin to differ. To ensure that certification programs overcome these challenges and address their focal problems as holistically as possible, their developers should draw on the expertise of a wide range of stakeholders and carefully reconcile these stakeholders’ differing perspectives.

\section{AI Certification Programs and Proposals}
\label{3}

To identify existing AI certification programs, we conducted internet searches, monitored social media, solicited input from colleagues, used our own prior knowledge, and examined references from documents about the programs and other relevant literature. These programs are not exhaustive, but are largely representative of the variation in AI-related certification.\footnote{ Additional certification programs under development include AI Global’s AI system certification \cite{57} CertNexus’s certified ethical emerging technologist program \cite{58} and ForHumanity’s compliance audit for AI systems and organizations \cite{86}. A number of universities offer AI-ethics related certifications or minors, including San Francisco State \cite{59} and Carnegie Mellon \cite{60}. Consultancies offer algorithmic auditing services, which may confer a certification \cite{61}.} We assessed the programs using publicly available documents. To fill in gaps in the documents, we conducted semi-structured interviews with representatives of certification program organizations. In total, we conducted three such interviews across seven identified AI certification programs:

\begin{enumerate}
	\item[A.] European Commission White Paper on Artificial Intelligence
	\item[B.] IEEE Ethics Certification Program for Autonomous and Intelligence Systems (ECPAIS)
	\item[C.] Malta’s AI Innovative Technology Arrangement (AI ITA)
	\item[D.] Turing Certification proposed by Australia’s Chief Scientist
	\item[E.] Queen's University’s Principles of AI Implementation executive education
	\item[F.] Finland’s civics course Elements of AI
	\item[G.] Danish labeling program for IT-security and responsible use of data
\end{enumerate}

The seven programs classify into four categories: self-certification of AI systems (A), third-party certification of AI systems (A, B, C, D), third-party certification of individuals (E, F), and third-party certification of organizations (D, G). Table \ref{table_2} presents our ratings of these categories in terms of feasibility, symbol-substance coupling, and means-ends coupling. The ratings are based on our analysis of the details of the seven programs (see below) in context of the management literature as reviewed above.

\begin{table}[b]
\caption{Assessment of Categories of AI Certification Programs}
\label{table_2}
\scriptsize
\begin{tabularx}{\textwidth}{ p{5.5cm} X X X X }
	\toprule
	\textbf{Type of Certification Program} & \textbf{Examples} & \textbf{Feasibility} & \textbf{Symbol-Substance Coupling} & \textbf{Means-Ends Coupling} \\
	\midrule
	Self-certification of AI systems & A & High & Low & Medium \\
	Third-party certification of AI systems & A, B, C, D & Medium & High & Medium \\
	Third-party certification of individuals & E, F & Medium & Low & Low \\
	Third-party certification of organizations & D, G & Medium & High & Medium \\
	\bottomrule
\end{tabularx}
\begin{tablenotes}
	\vspace{0.1cm}
	\item Ratings are authors’ judgments. Example letters are specific AI certification programs corresponding to the list in the opening of Section \ref{3} and the subsections of Section \ref{3}.
\end{tablenotes}
\end{table}

Self-certification is widely used in many industries globally, including suppliers’ declarations of conformity used in telecommunications and motor-vehicle manufacturing \cite{62}. For self-certification of AI systems, we rate certification programs in this category as having high feasibility because it requires little involvement from non-applicant actors and limited back-and-forth between them and the applicants. Symbol-substance coupling is low because of its inherent conflict of interest and lack of strong enforcement and assessment. These problems can be attenuated via strong ex-post enforcement after harms occur via liability regimes. However, liability is problematic especially for high-risk applications where the initial harm could be catastrophic, and because legal liability adjudication poses uncertainties for AI developers \cite{63}. Finally, means-ends coupling is medium because it encourages applicants to take ownership of the certification process and all that it represents, including by documenting their AI practices and thinking systemically about the corresponding ethical implications.

Third-party product certification is used in numerous industries globally, including Underwriters Laboratories certifications for electrical and fire safety and the Common Criteria \cite{64} for product cybersecurity certification. For third-party certification of AI systems, we rate certification programs in this category as having medium feasibility because AI systems are a relatively well-defined object to assess and certify, but they are also a novel one for which prior experience is of limited use. The need for qualified third-party organizations to offer assessment adds additional complications, including questions around costs and accountability (\cite{5} p.12). Symbol-substance coupling is medium because, on the one hand, third parties can assess certification claims and demand for certified systems could drive substantive adoption, but on the other hand, actual implementation can fall short on both meaningful assessment and enforcement. Finally, means-ends coupling is medium because of the potential for, but difficulty of, successfully assessing each AI system in its development and deployment context.

Third-party certification of individuals is common in a number of professions, including among engineers, lawyers, and actuaries, as well as programs like the Certified Information Privacy Professional credential \cite{65}. We rate AI-related certification programs in this category as having medium feasibility. A core variable is whether certification criteria include matters related to moral factors in addition to the usual criteria of technical knowledge and skills.\footnote{ The authors thank Jonathan Aikman for this point.} Assessment of moral standards is important but less feasible. Moral character is not readily testable, and education institutions may have less financial incentive to do so. Symbol-substance coupling is low because the substance of certification criteria are dynamic. Methods in explainability, fairness, safety, etc. are active research topics; certificates attesting to knowledge of them can become quickly outdated. This problem could be addressed via ongoing audits or education requirements comparable to those sometimes required for lawyers, but current AI certification programs lack such measures. Finally, means-ends coupling is low because, absent robust and dynamic moral standards, certification is unlikely to guarantee that the individuals will achieve sound performance on AI issues.

Third-party certification of organizations include the UK Cyber Essentials \cite{66} cybersecurity company certification and the FairTrade \cite{67} licensing contract. We rate AI-related certification programs in this category as having medium feasibility because it can draw on a preexisting body of experience and knowledge in corporate governance and related fields, but it still requires AI-specific innovations. Symbol-substance coupling is high because it can use robust preexisting methods like audits and data security management. Organizations are often familiar with these methods and understand how to respond with substantive activity. Their familiarity could also be used to game the system, though auditors are themselves familiar with such tactics. Finally, means-ends coupling is medium because assessments can support the adoption of a ‘systemic mindset’ but certification criteria may not be easily customized for the context of a particular organization.

\subsection{European Commission White Paper on AI}
\label{3.1}

The European Commission White Paper on AI \cite{10} proposes separate certification programs for low-risk and high-risk applications. Both programs would draw on prior certification programs, in particular those of the European Economic Area ``CE marking'' rules and the 2019 EU Cybersecurity Act. Both programs would also use similar certification criteria, such as the EC’s AI Ethics Guidelines for Trustworthy AI \cite{68}. If implemented, the programs would be important in their own right and could further be influential for AI certification beyond the single market, as is often the case for EU policies \cite{69}.

For low-risk applications, we interpret the White Paper as calling for voluntary self-certification. This is clear from the White Paper’s call for following the precedents of CE marking and the Cybersecurity Act, both of which use self-certification for low-risk applications. There is some ambiguity because the White Paper also suggests for the low-risk program to use a ``combination of ex ante and ex post enforcement'' (\cite{10} p.24). Ex ante enforcement would seem to preclude self-certification. We interpret a stronger significance to the two self-certification precedents and therefore believe that the White Paper intends the low-risk program to involve voluntary self-certification. AI applicants could attest to their system having met the criteria, and face ex post penalties if their system falls short. Applicant participation would be incentivized by consumer demand for certificates of AI trustworthiness \cite{10}.

To be effective, the low-risk program would need to be carefully designed. First, it needs an enforcement mechanism with clear, strong penalties for applicants who falsely certify their AI systems as meeting the specified criteria (\cite{70} p.5). Otherwise, there could be symbol-substance decoupling, with applicants rubber-stamping their systems as meeting the criteria. Fortunately, the program could build on the compliance infrastructure that already exists for CE markings. Second, the criteria need to capture the many aspects of AI trustworthiness as defined in the ``Ethics guidelines for trustworthy AI'' \cite{68}, which include, among other things, fairness, security, and accountability. The criteria further must promote the systemic mindset needed to advance this conception of trustworthiness. Otherwise, there could be means-ends decoupling, with the certification program not advancing its underlying goals.

For high-risk applications, the White Paper clearly specifies mandatory third-party certification via what it refers to as ``conformity assessment'' (\cite{10} p.23). Its procedures would be based on CE marking rules or the Cybersecurity Act \cite{10}, both of which require third-party verification for high-risk applications. The White Paper’s program would use existing accredited auditing organizations (``notified bodies'') of the Member States (\cite{10} p.25). EU regulation would compel AI applicant adoption and compliance.

To be effective, the high-risk program will also need several design elements. First, it needs to ensure that the certifying bodies have the technical expertise and capacity needed to assess AI systems. Using the existing notified bodies makes the program easier to implement, but the bodies currently lack expertise and capacity on AI. Unless this situation is improved, the assessments could be inaccurate and AI applicants may be able to obtain symbolic certification. Additionally, as with the low-risk program, the criteria must be carefully specified to ensure means-ends coupling.

\subsection{IEEE Ethics Certification Program for Autonomous and Intelligent Systems (ECPAIS)}
\label{3.2}

The IEEE ECPAIS program \cite{11} is currently under development. It considers driving factors and inhibitors of AI system transparency, accountability, and bias. Offering a series of recommendations, the program is expected to yield a quality mark. The developer of an AI system can use the quality mark in presenting the system if it follows the recommendations in a checklist fashion. The program is expected to be administered by an organization other than the IEEE, which will assess certification applications. Whether the program will use an auditor to verify these applications has yet to be determined. The program is voluntary, with expected consumer demand as the driver for adoption.

To be effective, EPCAIS would likely need the following. First, it would need to incentivize substantive adoption by generating consumer awareness and demand or by establishing an auditing requirement. Second, to improve means-ends coupling, it should offer distinct quality marks for transparency, accountability, and bias; it should also promote a systems mindset and consider the context in which the system is developed or deployed. Third, it should develop in-house expertise on certification; as far as the authors know, the IEEE has not launched a system-level certification program before. Finally, it should foster an enthusiastic audience in order to successfully launch. However, given that ECPAIS remains under development, we cannot offer a conclusive assessment.

\subsection{Malta AI Innovative Technology Arrangement (AI ITA)}
\label{3.3}

The Maltese government launched the AI ITA certification program \cite{13} in summer 2020. The program certifies AI systems to a particular use case following review of an application detailing development and maintenance processes, corporate governance structures, and a responsible administrator. Applications must describe how the system handles failure modes and conforms to Malta’s ethical AI framework, among other requirements \cite{13}. The program also (in most cases) requires applicant systems to have two software features: a ``harness'' that safeguards how the system handles specified failure modes and a “forensic node” that logs activity and is stored physically within Maltese jurisdiction. The applicant system must also disclose specific elements of its compliance with the ethical AI framework to end-users via terms of service. Applications are reviewed by the Malta Digital Innovation Authority (MDIA) and are subsequently confirmed by an accredited auditor. The program is voluntary, with adoption expected to be driven by demand for certified systems from government procurers, businesses, and consumers.

The Malta program has high feasibility because it benefits from a dedicated regulator which was established to manage this and similar programs. This dedicated expertise in assessing certification as well as accrediting third-party auditors can reduce concerns of symbol-substance decoupling. The program’s focus on awarding certification for a specific system in a specific use case based on an assessment that includes the organization and a responsible individual incorporates a wide context that can help reduce concerns of means-ends decoupling. However, no application has yet been submitted. The effectiveness of the software harness to mitigate failure modes and the usefulness of mandated ethics disclosures in terms of service remain to be seen.

\subsection{Australian Chief Scientist Proposal: Turing Certification}
\label{3.4}

The Turing Certification program was proposed by Australia’s Chief Scientist Alan Finkel in 2018 \cite{12,71}. It would certify AI developer organizations and low-risk AI systems, drawing inspiration from FairTrade certification among other programs \cite{72}. This dual certification aspect could be especially impactful: it promotes a wide scope of assessment, thus improving means-ends coupling, and its organizational audit further improves symbol-substance coupling. As proposed, Turing Certification would be a voluntary program with third-party certifiers, with participation incentivized by demand from purchasers including consumers and governments. Third-party audit would confirm that the applying company, its development processes, and end AI system conform to not-yet-set standards for trustworthy AI. The proposal has not been enacted; it is unclear if it ever will be. Its most recent public discussion was in 2019 (\cite{73} p.161). If it is to be enacted, it would need clear certification criteria, a substantial challenge given the lack of agreed-upon trustworthy AI standards today.

\subsection{Queen’s University Executive Education: Principles of AI Implementation}
\label{3.5}

In partnership with IEEE, Queen’s University has launched an executive education certificate program on Principles of AI Implementation \cite{14}. Demand is likely to be driven by employers that value certified employees. The program has not yet run, and it is unclear how strong employer demand will be. The program covers ethical design principles, considerations for building trust in AI applications, and current Canadian regulatory requirements. This broad curriculum could support a ``systemic mindset'', thereby improving means-ends coupling. However, these are dynamic topics; the program will need to stay up to date to retain means-ends coupling. Additionally, it is only a two-day program; this short duration precludes much substantive learning, creating risk of symbol-substance decoupling.

\subsection{Finland Civics Course: Elements of AI}
\label{3.6}

The government of Finland, in a partnership between the University of Helsinki and the consultancy Reaktor, offers certificates to people who complete its civic education course Elements of AI. Certificates are awarded for all who complete at least 90\% of the exercise and score 50\% or above on attempted exercise questions \cite{74}. Credits can be transferred to the University of Helsinki. Demand may be driven by individuals’ curiosity and sense of civic responsibility. It is unclear if employers would be interested in hiring certified individuals, and its introductory content would be insufficient for professional education, though that is not the program’s aim. Thus far, demand has been significant: over 500,000 people have signed up and students from 170 countries have completed the course \cite{15}.

\subsection{Danish labeling program for IT-security and responsible use of data}
\label{3.7}

The Confederation of Danish Industry, Danish Chamber of Commerce, SMEdenmark, and Danish Consumer Council \cite{16} are currently developing a certification program for information technology security and data ethics. The program does not primarily focus on AI, but it was listed by the European Commission (\cite{10} p.10) as an AI activity, and an interview and review of internal documents confirms that AI is within its scope. The program scope is ambitious, covering elements ranging from IT security to data management, to processes to address algorithmic bias. This broad scope could support a ``systemic mindset'' in the organization, and thus improve means-end coupling. The same scope may undermine symbol-substance coupling, however, as it could be difficult to achieve compliance with and effectively communicate to external stakeholders about so many certification criteria. Eventual administration of the program could encounter a challenge in moral hazard because a single organization—funded by certification fees—sets certification criteria, reviews applications, and performs (limited) audits.

\section{Certification for Future AI Technology}
\label{4}

An essential attribute of AI technology is that it continues to evolve. Certification programs designed for the AI of today may not fare well with the AI of tomorrow. Adjustments will be needed to keep the programs up to date. However, adjustments can be expensive and are therefore not always made. Where possible, current certification programs should be designed to accommodate potential future directions in AI. This is no easy task: the future of AI cannot readily be predicted. Many have tried and failed \cite{75}. Nonetheless, some broad contours of future AI can be at least tentatively described, enough to derive some implications for certification programs. Toward the end of this section, we offer some remarks specific to the most advanced long-term AI, though much of the discussion may be more applicable to AI that will be developed in the interim, i.e., medium-term AI \cite{76}.

Some elements of AI certification are relatively durable in the face of changes in the technology. First, many of the underlying goals for certification programs, such as fairness and beneficence, derive from universal ethics principles that apply not just to any AI technology, but to society in general. These ethics principles are used as a ``North Star'' guide in today’s certification programs (\cite{36} p.35) and will likely remain relevant as the technology matures. Future programs might potentially even certify the ethics of either the AI system or its developers. Second, some aspects of certification are matched to the types of people and institutions that are involved in AI, such as business executives, citizens, and corporations. It is fair to expect that such actors will remain involved in AI over the years. Certification program capacity to engage these actors may likewise remain relevant. That includes matters such as the capacity to improve corporate transparency and to educate large numbers of citizens.

Similarly, one way for certification programs to remain relevant over time is to emphasize human and institutional factors. As discussed in Section II, certification programs can improve symbol-substance and means-ends coupling by promoting certain values and norms among applicants. Among other things, doing so positions applicants to act ethically as new challenges arise. Today, we do not know exactly what new challenges will arise for future AI, but we can be very confident that there will be new challenges. Certification programs should endeavor to make it more likely that whoever faces these new challenges will act according to a high ethical standard.

Certification programs can also remain relevant by building-in mechanisms for updating their certification criteria. For example, in the establishment of a new AI certification program, it could be specified that an expert body meets periodically to review and update the criteria, as is common with technical standards bodies. Programs could also include sunset provisions such that a program will cease to exist unless its criteria are updated. Additionally, certification programs could include a budget for ongoing research so that knowledge about how to update the criteria is available when needed.

The field of AI research could also be a focus of certification. Future AI techniques will derive from current or future research. Some techniques may be better able to meet ethics standards; these techniques could be an object of certification. Academia already has institutional review boards to certify human subjects research as meeting standards for responsible conduct of research. Similar boards have been proposed for overseeing future AI research \cite{77}. Importantly, such boards should focus not just on whether research meets procedural standards such as harm to research subjects, but also on whether it meets standards in terms of expected impacts on broader society \cite{78}. However, it can be difficult to know in advance which techniques will perform well—hence the need to do the research. Therefore, research is another domain in which certification may do well to emphasize human and institutional factors.

Some forms of future AI may not be conducive to certification. In particular, AI systems that reach or exceed human-level intelligence in many domains \cite{79,80} might not be a viable object of certification because humans could be powerless to take corrective actions suggested by the certification. To the extent that such AI is a significant concern, certification may only play a role in its development, while humans are still in command. This is one area in which certification of AI research programs may be especially significant, such as to promote trust among rival AI development groups \cite{81} or share information about safety and ethics measures to alleviate rivals’ concerns \cite{82}. Additionally, if the AI is developed by national government—a distinct possibility given its national security implications—then there may be a role for an international certification regime, perhaps analogous to monitoring and verification regimes that have been established in other domains such as for biological, chemical, and nuclear materials. Should humanity ever face the prospect of developing such a technology, it would probably want to know that it was being developed according to a high ethical standard. That could be an important role for future AI certification programs.

\section{Conclusion}
\label{5}

Certification is a valuable tool for addressing information asymmetries and incentivizing better behavior. Certification can provide information about AI systems themselves as well as the organizations and individuals that are involved with them. AI-related programs could include both voluntary certification to ethics principles and mandatory conformity assessment to regulatory requirements. Today, a variety of AI certification programs are in development and use. These programs are focused on current AI technology, but certification can also play a role in future AI systems, up to including the most advanced long-term AI.

One gap in the current AI certification landscape is the processes through which AI is developed. Although current certification programs assess the AI development process to varying degrees, no program certifies the process as its object. As a consequence, certification programs may fail to address process-oriented Potemkin attributes with maximal symbol-substance coupling and may fail to promote constructive organizational processes that can improve means-ends coupling. Improved attention to processes as the focus of certification could draw on, among other things, recent work on algorithmic auditing \cite{36} and existing programs such as ISO 9001 on quality management \cite{25} and ISO 27001 on information security management \cite{83}. Early work at ISO on an AI management system standard may one day yield such a process-focused AI certification program \cite{84}.

Another challenge is to improve the customization of AI certification for particular circumstances. Whereas some, such as the Malta AI ITA, are customized for specific jurisdictions, others, such as IEEE ECPAIS, make homogeneous prescriptions for AI developed worldwide. Additionally, most of the certification programs lack customization for each of the diverse sectors to which AI is applied, e.g., healthcare, transportation, and national security. Similarly, the programs tend to treat ``AI'' as a monolithic technology instead of being customized for the diverse types of AI that exist, e.g., language translation, facial recognition, and anomaly detection. Homogenous certification may similarly encounter challenges when an AI system is but a small, inextricable component of a larger digital system or workflow. All of this is a problem because certification programs that focus on specific niches tend to have better means-ends coupling \cite{21}.

Many of the AI certification programs discussed here rely on, or at least strongly benefit from, demand from the customers of AI systems, including consumers, businesses, and governments. This is especially important for voluntary programs, where demand usually is the primary driver of participation. For good results to accrue, customers must not only demand certification—they must demand good certification. To that end, customers must be sufficiently educated about the technology and the associated ethics and safety concerns (\cite{68} p.23). Education programs such as Finland’s Elements of AI civics course could play an important role.

Overall, AI certification remains at an early stage. Much remains to be seen, including which certification programs will be implemented, how effective they will be, how difficult or expensive they will be to implement, and how durable they will be in the face of changes in AI technology. All these aspects will influence the likelihood for adoption \cite{85} and their success at symbol-substance and means-ends coupling. This paper has presented a snapshot in time, with an assessment of an indicative sample of AI certification programs. It has also identified key concepts within the management literature on certification. Both contributions should inform future monitoring and evaluation of AI certification as the field matures. Future research would do well to track which certification programs are performing well, to learn from their successes and failures, and to identify gaps in the field and areas for improvement. It will likewise be important to monitor changes in AI technology and to ensure that certification programs stay up to date. AI certification is not a panacea, but it can play a valuable role in the overall mix of AI governance tools.

\section*{Acknowledgements}

This paper does not reflect the views of the authors’ employers. The authors thank David Danks, Laura Galindo-Romero, Mikael Jensen, Dewey Murdick, Cullen O’Keefe, and Nell Watson, and participants of the workshop on AI and Soft Law at Arizona State University for their insightful feedback on an earlier version of this paper. The authors also thank Trevor Sammut and two other interviewees for their insights. All errors are the authors’ alone.

\bibliographystyle{unsrt}
\bibliography{ms}

\end{document}